\documentclass{camera}
\usepackage{graphicx}  

\begin{document}

%
\title{The beginning of cosmic ray astronomy}

%
\author{Todor Stanev}

%
\organization{Bartol Research Institute, Department of physics and Astronomy,
 University of Delaware, Newark, DE 19716, U.S.A.}

\maketitle

\begin{abstract}
 We discuss the anisotropic arrival directions of the ultra high 
 energy cosmic rays detected by Auger which I consider one
 of the biggest discoverie in astrophysics during the last year.
\end{abstract}

%

\section{Introduction}

 As you may conclude from the different concluding remarks at
 this meeting there are various approaches to doing them.
 Most of my esteemed colleagues spoke on several topic that
 they consider important now and in the future. 
 I decided to concentrate on one single topic that excited 
 not only me but a large number of astrophysicists that were
 never interested in ultra high energy cosmic rays (UHECR) -
 the observed correlation of the highest energy events detected
 by the Pierre Auger Observatory in Argentina with active
 galactic nuclei (AGN)~\cite{Auger1,Auger2}. This discovery 
 marked the beginning of a new type of astronomy - cosmic ray 
 astronomy - that may become as important in the future as 
 the TeV gamma ray astronomy has recently become. 

 TeV gamma ray astronomy does not only reveal the type of 
 the gamma ray sources, it also provides an indirect measurement
 of the infrared/optical background whose estimates are based on
 theoretical calculations involving the light emission of different
 types of stars and galaxies and the cosmological evolution of
 these objects and the matter density around them in the Universe.
 Cosmic ray astronomy also involves some general features of the
 nearby Universe that are otherwise extremely difficult to 
 measure such as the local (within 200 Mpc) distribution of matter and
 different powerful astrophysical objects, the average strength of
 the intergalactic magnetic fields (that could even be mapped when
 the sources are confirmed) and the galactic ones.

 I will discuss these results and report on the small contribution
 that I have made to this topic. It is not connected to the type 
 of the UHECR sources, but includes a better than 3$\sigma$
 confirmation of the anisotropy detected by Auger~\cite{Stanev08}.

\section{The anisotropy discovered by Auger}

   The Auger Collaboration reported a correlation of their
 27 highest energy events (E $>$ 5.7$\times$10$^{19}$ eV (57 EeV)
 with active galactic nuclei (AGN) at redshifts $z$ less than
 0.017~\cite{Auger1,Auger2} from the V\`{e}ron-Cetty and V\`{e}ron
 catalog~\cite{Veron} (V-C). Twenty out of 27 events are within 3.1$^o$
 of individual AGN, while for
 isotropic arrival distribution one expects on average 5 coincidences.
 Five of the non-correlating events come from less than 12$^o$
 galactic latitude which may be understood as larger deflections in
 the galactic magnetic field. The scattering in the galactic magnetic
 field is consisten with these events being protons, not heavier
 nuclei. It is surprising
 that there are no events coming from the direction of the Virgo
 cluster, that includes a large number of powerful galaxies in
 addition to M87, as stated in Ref.~\cite{Gorbunovetal08}.

 A better description of the statistical analysis and a broad discussion
 if its features is given in the longer second paper~\cite{Auger2}.
 The self correlation between the 27 events was studied and found to
 be consistent with the existence of many (more than 61) sources 
 within the 71 Mpc radius, that leads to a source density 
 of 3.5$\times$10$^{-5}$ Mpc$^{-3}$. The Auger Collaboration warns us
 that because most of the correlating AGN are not very powerful the
 observed anisotropy may be due to a different type of source.

 It is quite interesting that two Auger events coincide with the
 close by (less than 4 Mpc) powerful radio galaxy Cen A, and several
 other events are close to it.

 Among the criticism toward this analysis is the statement that
 redshift $z$ of 0.017 is much lower than the expected contribution
 of redshifts for extragalactic protons and nuclei. Protons are supposed
 to come
 to us from redshifts up to 0.05, roughly distances to 200 Mpc.
 Other authors complain about the types of AGN in the
 V-C catalog that are low luminosity Seyfert and LINER galaxies.
 Since I do not have much knowledge about AGN I
 decided to check the discovered anisotropy by studying the
 correlation of the Auger events with the Supergalactic 
 plane (SGP).

\section{Correlation of the Auger events with the Supergalactic plane}

 A report on the correlation of the then existing statistics 
 of UHECR of energy above 4$\times$10$^{19}$ eV with the
 Supergalactic plane~\cite{deV1} was published
 in 1995~\cite{Stanevetal95}. The data set consisted mostly
 of events detected by the  Haverah Park detector with the addition
 of events  detected by AGASA, Yakutsk and Volcano Ranch air
 shower arrays.
 The anisotropy of that data set was studied by a comparison of 
 the average and RMS supergalactic latitudes $|b_{SGP}|$ of the
 experimental events to that of an isotropic Monte Carlo sample.
 The significance of the correlation was at the 3$\sigma$ level.
 Later on, when the AGASA detector dominated the
 world's statistics, the  significance of the
 correlation decreased.
\begin{figure}[thb] 
\centerline{\includegraphics[width=288pt]{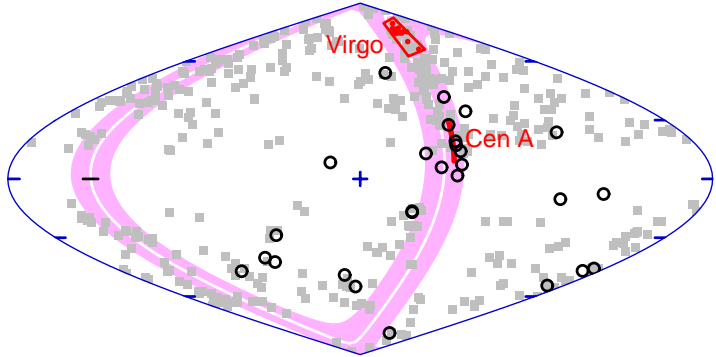}}
\caption{The 27 Auger events are superimposed
 in galactic coordinates in a sinusoidal projection on the 
 supergalactic plane in the definition of Ref.\cite{deV1}.
 The shaded area shows $|b_{SGP}| <$ 10$^o$. The gray squares
 show the positions of V-C AGN that have galactic latitude 
 more than 12$^o$. The Virgo cluster and Cen A are also shown.
\label{auger17}
}
\end{figure}
 We decided to check on the correlation of the 27 Auger events
 with the same definition of the nearby large scale mass 
 distribution. The result is shown in Fig.~\ref{auger17}.
 The probability for such a correlation was calculated 
 with a Monte Carlo simulation where 100,000 sets of 27 events
 were injected isotropically in the Auger field of view. 
 We arbitrarily  chose to bands of 10$^o$ and 15$^o$ in 
 supergalactic latitude that surround the 1$\sigma$ deviation
 of the AGN from the V-C catalog to the SGP. 
 Nine (13) of the 27 experimental events fall in these two
 bands. The random coincidence probability derived from the
 Monte Carlo is 0.024 (0.008) corresponding to
 2.0 (2.4) $\sigma$.

\begin{figure}[thb] 
\centerline{\includegraphics[width=288pt]{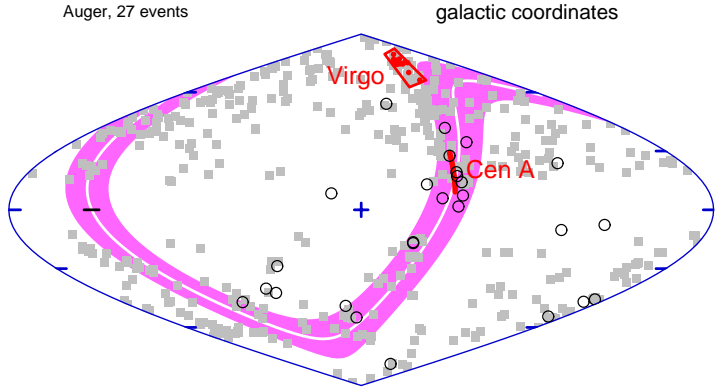}}
\caption{The supergalactic plane corrected as in Ref.~\cite{Lahavetal}
 for distances up to 70 Mpc.   
\label{auger16}
}
\end{figure}

 Until now we dealt with the original definition~\cite{deV1}
 of the supergalactic plane by de Vaucouleurs. This definition was
 studied in 2000 by Lahav et al. in terms of redshift~\cite{Lahavetal}
 on the basis of the Optical Redshift Survey (ORS)
 (8457 galaxies, 98\% with redshift). Since ORS had a zone of avoidance
 of $|b| <$ 20$^o$ it was complemented with IRAS galaxies in order
 to describe better the intersection of SGP with the galactic plane.   
 This study introduces correction to the definition of SGP for 
 different distances from 0 to 80 Mpc. The one that is most suitable
 for analysis  of the Auger events is for distances up to 70 Mpc,
 i.e. identical to the redshift of 0.017.
 The corrected shape of the SGP is plotted in Fig.~\ref{auger16} the same
 way as the original definition is plotted  in Fig.~\ref{auger17}.

  One can now count how many UHECR are in the 
 10$^o$ and 15$^o$ band with the new definition of the
 supergalactic plane. The numbers are  13 and 15 events
 respectively. The probability to have so many 
 events from isotropic arrival direction distribution
 drops to 1.6$\times$10$^{-4}$ (6.0$\times$10$^{-4}$) 
 for distances from the supergalactic plane of 10$^o$(15$^o$).
 The correlation of the Auger events with the updated definition
 of the SGP within 70 Mpc is indeed much more significant - the
 probabilities above correspond to 3.6(3.2)$\sigma$. 
 Since the rotation angle of SGP in galactic coordinates in
 the analysis of Lahav et al. depends on the the redshifts
 of the astrophysical objects the one we used certainly defines
 better the plane of weight of the mass distribution within
 70 Mpc and of the UHECR sources if these cosmic rays indeed
 are produced within this distance. Even if the GZK horizon is
 larger we expect some degree of correlation with the nearby 
 astrophysical objects.
 
 It is important to note that the HiRes Collaboration does not
 confirm the correlation with AGN~\cite{HiRes_AGN}. Out of 13
 highest energy events only two correlate with AGN from the 
 V-C catalog within 3.1$^o$. The HiRes paper does not describe
 well their field of view and it is difficult to understand if
 the difference in the field of view is the reason for this
 discreapancy. There is also a lack of obvious correlation 
 with SGP. Three (5) of the HiRes events are within 10$^o$
 (15$^o$) of the SGP in the definition of Ref.~\cite{Lahavetal}.
 
 We still believe that the anisotropy published by the Auger
 Collaboration is strong and with the increase of the 
 experimental statistics, which must have doubled by August 2008, 
 it will lead us to identification 
 of the sources of the highest energy cosmic rays.



%

\begin{thebibliography}{99}
\bibitem{Auger1} The Auger Collaboration (J.~Abraham et al.), Science
 318:938 (2007)
\vspace*{-3truemm}
\bibitem{Auger2} The Auger Collaboration (J.~Abraham et al.), Astropart. Phys.
29:188 (2008)
\vspace*{-3truemm}
\bibitem{Stanev08} T.~Stanev, arXiv:0805.1746
\vspace*{-3truemm}
\bibitem{Veron} M.-P.~V\'{e}ron-Cetty \& P.~V\'{e}ron, A\&A {\bf 455},
 773 (2006)
\vspace*{-3truemm}
\bibitem{Gorbunovetal08} D.S. Gorbunov et al., arXiv:0804.1088 (2008)
\vspace*{-3truemm}
\bibitem{deV1} G.~de Vaucouleurs et al., The Third Catalogue of Bright
Galaxies (RC3) University of Texas Press, Austin (1991)
\vspace*{-3truemm}
\bibitem{Stanevetal95} T.~Stanev et al., Phys. Rev. Lett., {\bf 75}:305 (1995)
\vspace*{-3truemm}
\bibitem{Lahavetal} O.~Lahav et al., MNRAS, {\bf 312} 166 (2000)
\vspace*{-3truemm}
\bibitem{HiRes_AGN}  HiRes Collaboration (R.U.~Abbasi et al), arXiv:0804.0382 

\end{thebibliography}
\end{document}